# Multistage smoothing based multistep pulse compressor for ultrahigh peak power lasers

SHUMAN DU[1, 2, 3], XIONG SHEN[1, 3], WENHAI LIANG[1, 2], PENG WANG[1], JUN LIU[1, 2, *], RUXIN LI[1, 2]

[1] *State Key Laboratory of High Field Laser Physics and CAS Center for Excellence in Ultra-intense Laser Science, Shanghai Institute of Optics and Fine Mechanics, Chinese Academy of Sciences, Shanghai 201800, China*

[2] *University Center of Materials Science and Optoelectronics Engineering, University of Chinese Academy of Sciences, Beijing 100049, China*

[3] *These authors contributed equally to this work.*

*\* jliu@siom.ac.cn*

**Abstract:** Ultrahigh peak power lasers are important scientific tools for frontier laser-physics researches, in which both the peak power improvement and operating safety are very important meanwhile limited by the damage threshold and size of compression gratings currently. Based on a recent reported method "multistep pulse compressor (MPC)", a multistage smoothing based MPC (MS-MPC) is proposed here to further improve the running safety, operating convenience, and simplify the whole setup of the MPC. In this optimized design, the beam smoothing is not simply executed in the pre-compressor or main-compressor, but separated into multistage. Then, it can protect important optics in every stage directly and reduce the executing difficult of typical MPC at the same time. The prism pair based pre-compressor will induce suitable spatial dispersion which is easier to be achieved and enough to protect the first grating directly. At the same time, the asymmetric four-grating compressor (AFGC) will also induce spatial dispersion to further smooth the laser beam which helps to protect the last grating directly. In this way, 10s-100s PW lasers can be compressed by using current available optics with improved operating safety owing to remove random spatial intensity modulations. Furthermore, an additional beam smoothing stage can be added before the main amplifier to protect the biggest amplification crystal away from damage. This MS-MPC optical design can be easily extended to be used in all exist PW laser facilities to improve their potential compressed pulse energy and running safety.

## 1. Introduction

Owing to the great improvement of chirped pulse amplification (CPA) and optical parametric chirped pulse amplification (OPCPA) techniques [1, 2], ultrahigh peak power lasers have obtained great progress in recent years. So far, nearly fifty petawatt (PW) level laser facilities have been constructed around the world [3]. These ultrahigh peak power lasers have important applications in many frontier scientific researches such as laser particle accelerator and strong field quantum electric dynamic. In very recent, 10 PW lasers had been achieved on both SULF and ELI facilities from China and Europe, respectively [4, 5]. Furthermore, 10s to 100s PW lasers have been proposed based on OPCPA on several laser facilities, such as OPAL-75PW (United States) [6], XCELS-200PW (Russian) [7], ELI-200PW (Europe) [8], and SEL-100PW (China) facilities [9], among them the SEL-100PW laser facility is in constructing right now which is expected to reach 100 PW in the next few years.

In a PW laser system based on either CPA or OPCPA method, a stretcher is added before the amplifier to broaden the laser pulse duration from femtosecond scale to nanosecond level to avoid the laser induced damage on crystals, accordingly a pulse compressor is added after the amplifier to compress the chirped nanosecond laser pulse back to femtosecond so as to achieve ultrahigh peak power output. Basically, in both CPA and OPCPA methods, the damage problem of the amplification crystals due to the limited damage threshold and confining size of crystals is transferred to the pulse duration in time domain, a stretcher and a compressor are added before and after the amplifier, respectively, to solve the induced temporal problem specifically. When the laser peak power reach to 10s to 100s PW, new problem appears in the pulse compressor this time, due to the damage problem of compression grating because of the limited damage threshold and size again. To solve this problem, previously proposed optical designs of XCELS-200PW, ELI-200PW, and SEL-100PW were all based on the multi-beam combination method named multi-beam tiled-aperture combing [10], which is very complicated and difficult because beam combing is very sensitive to the differences of optical delay, pointing stability, beam wavefront, and dispersion among every beams [11, 12]. Recently, a novel method named multistep pulse compressor (MPC) has been proposed to solve the compression problem, in which the limited input/output pulse energy problem is transferred to the spatiotemporal properties of the input/output laser beam [13]. The same as CPA or OPCPA, a pre-compressor and a post-compressor are added before and after a traditional Treacy four-grating compressor (FGC) to solve the induced spatiotemporal problems specifically. Furthermore, as a simplified optical design of the MPC, asymmetric FGC (AFGC) is proposed to replace both the prism pair based pre-compressor and the symmetric FGC simultaneously in a MPC [14]. Then, the typical three-step MPC is reduced to two-step MPC with the AFGC design. This AFGC can be easily modified in almost all previous PW laser facilities with typical symmetric FGC, and improve their operating safety and maximum bearable output pulse energy.

In this paper, we combine both the typical three-step MPC and the AFGC together in the pulse compressor, which is named as multistage smoothing based MPC (MS-MPC). In the optical design of the MS-MPC, the symmetric FGC will be replaced by an AFGC used as the main compressor. This combination owes several obvious advantages in comparison to single typical three-step MPC or single AFGC based two-step MPC optical design. Firstly, the pre-compressor with prism pair before the FGC can induce relative small suitable spatial dispersion width which is enough to smooth the laser beam for the safety of the first grating. Secondly, the AFGC will also induce suitable spatial dispersion width to protect the last grating directly, which is the weakest optics in the FGC. Thirdly, since it does not need very large spatial dispersion width in the pre-compressor which needs tens meters long distance by using prism pair with small apex angle, it is then easy and convenient to be achieved with saved optical space. The grating pair distance difference is very small that is also very easy to be obtained. Fourthly, every grating in FGC is well protected because spatial dispersion can be induced directly in every stage. Then, even if there is defect/damage appeared on one grating, this damage will not transfer to the next grating owing to the beam smoothing using spatial dispersion in every stage. While in traditional symmetric FGC, damage defects in one grating may cause hot spots due to diffraction, which can be transported to the last grating and damage it simultaneously. Furthermore, except for the compressor, an additional prism/grating pair based beam smoothing stage can be added before the main amplifier to smooth the input laser beam and protect the expensive and big amplification crystal.

## 2. Beam smoothing with spatial dispersion

2.1 Spatial intensity smoothing owing to spatial dispersion

Beam smoothing based on the induced spatial dispersion is one of the key effect that can reduce the laser spatial intensity modulation, and then improve the input/output laser pulse energy in the compressor stage of a PW laser system. So far, there are two typical optical designs, prism pair or grating pair, which can induce spatial dispersion to smooth the laser beam [13, 14]. In both setups, the first prism or grating will induce angular dispersion to the laser beam, the second parallel prism or grating is used to collimate the output laser beam. The induced spatial dispersion width $D$ is related to the laser spectral bandwidth and the distance between the prism pair or the grating pair. The induced spatial dispersion width $D$ can be expressed as $D=L(\tan(\theta_s)-\tan(\theta_l))\cos\alpha$. Where, $L$ is the perpendicular distance of the prism pair or grating pair, $\theta_s$ and $\theta_l$ are the diffraction angles of the shortest wavelength $\lambda_s$ and the longest wavelength $\lambda_l$, respectively, of the input laser pulse, $\alpha$ is the incident angle of the grating or the apex angle of the prism. For laser pulse with a narrowband spectrum, the difference of $\tan(\theta_s)-\tan(\theta_l)$ is small, then it needs much longer distance $L$ to achieve same amount of spatial dispersion width $D$ in comparison to that with a broadband spectrum. In comparison to prism pair, the grating pair can induce relative large spatial dispersion width with a short distance $L$ usually. Then, a transmitted grating pair can be used in the pre-compressor for laser pulses with narrowband spectrum, for example, picosecond PW system. As for an AFGC, it equals to a symmetric FGC combined with a grating pair with certain distance $d$ which is equal to the difference of the two grating pair distances of the AFGC. As a result, AFGC can also induce spatial dispersion the same as grating pair. The basic optical scheme of prism pair and grating pair for beam smoothing are shown in Fig. 1(a) and Fig. 1(b), respectively.

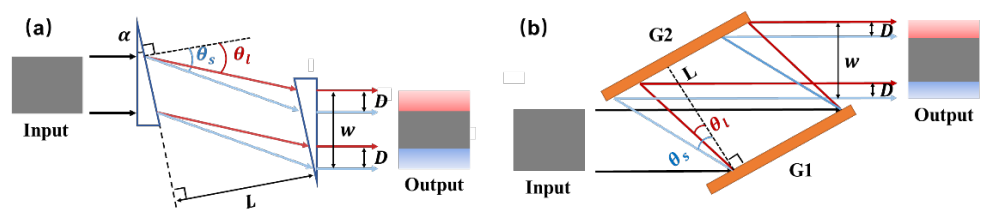

Fig.1. Optical scheme of beam smoothing based on prism pair (a) and grating pair (b). $\theta_s$ and $\theta_l$ are output angles of the shortest wavelength $\lambda_s$ and the longest wavelength $\lambda_l$, respectively, from the first prism or grating of the input laser pulse. $L$ is the distance of the prism pair or the grating pair. $D$ is the introduced spatial dispersion width on the output beam. $w$ is the full width of the output beam.

The beam smoothing effect in one direction is related to the induced spatial dispersion width $D$ directly. For both prism pair and grating pair, the same smoothing effect will be achieved if they induce same amount of spatial dispersion width $D$. Due to the relative weak angular dispersion ability, for a laser pulse at 925 nm central wavelength with 200 nm spectral bandwidth, it needs nearly 70 meters distance between the two fused silica prisms with 15 degrees apex angle to induce about 60 mm spatial dispersion width in the pre-compressor [14]. Prism pair with large apex angle will shorten the distance. However, the prism thickness with large apex angle will be increased, which will be more expensive and induce higher absorption energy. Two perpendicular prism pairs at both X and Y directions will shorten the distance which will be discussed in the following section. Transmitting grating may help to reduce the distance with high efficiency if it is OK in the future. In an AFGC, it is equal to a grating pair with reflective gratings. Since gratings own strong angular dispersion ability, for the same laser pulse at 925 nm with 200 nm spectral bandwidth, a grating pair with 1400 lines/mm groove density and around 60 degrees incident angle can induce more than 400 mm spatial dispersion width on gratings with only about 1.2 meters separated distance. As a result, the AFGC can also induce suitable spatial dispersion width with relative short different separated distances between the two grating pairs, which will further smooth the output laser beam. Since the AFGC can also smooth the laser beam, the pre-compressor with prism pair can induce a relative small amount of spatial dispersion width to the laser beam by using prism pairs with small apex angle, which is more feasible in a real laser system.

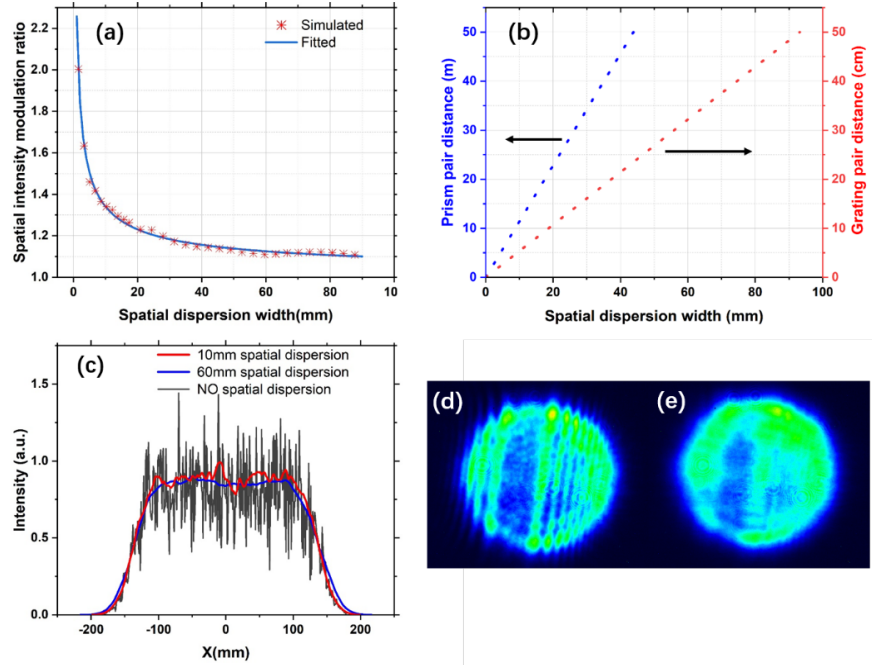

Fig 2. (a) Spatial intensity modulation ratio variation related to spatial dispersion width, (b) The spatial dispersion width changing curves with distance of prism pair or distance of grating, (c) Intensity curves of the center lines along the X axis of the beams without spatial dispersion (black) and those with 10 mm (red) or 60 mm (blue) spatial dispersion widths. (d) the output laser beam compressed by a typical symmetrical FGC, (e) the output laser beam compressed by AFGC.

To explain the beam smoothing capability based on induced spatial dispersion clearly, spatial intensity modulation with random high spatial frequency of 1.0 mm$^{-1}$ is intentionally induced into a laser beam before the prism pair or grating pair, which may due to defects in optics or dusts in optics and optical paths. Absolutely beam smoothing will be obtained for the output beam after either prism pair or grating pair, as shown in Fig. 2. A pulse centered at 925 nm with 200 nm full spectral bandwidth from an amplifier with 10$^{th}$ order super-Gaussian profile 370×370 mm$^2$ beam profile is used as the simulation laser source. As shown in Fig. 2(a), spatial intensity modulation experiences a quick decreasing from about 2.0 to 1.3 with spatial dispersion width increasing from zero to about 10 mm, and a mild decreasing to about 1.1 with spatial dispersion width continue increasing to about 60 mm, and then it can hardly be decreased more. Fig. 2(b) show the perpendicular distance of the prism pair with 15º apex angle or that of the grating pair with 1400 ln/mm corresponding to the induced spatial dispersion width. As we can see that the separated distance for prism pair is about tens of times longer than that of grating pair for achieving the same amount of *D*. About 8 mm spatial dispersion width can be obtained using the prism pair separated by about 10 m. Fig. 2(c) shows the intensity curves of the center lines along the X axis of the original 370×370 mm$^2$ beam with no spatial dispersion and that when 10 mm or 60 mm spatial dispersion width are introduced. We can see the strong spatial intensity modulation in the laser beam when there is no spatial dispersion. If a 10 mm or 60 mm spatial dispersion are introduced, the beam are both well smoothed, and its spatial intensity modulation is reduced to about 1.3 or 1.1, respectively. Note that the induced spatial intensity modulation with high spatial frequency is random, then there usually are spatial intensity modulation with low spatial frequency in the simulation, which cannot be smoothed. In reality, hot spots usually occurred sparsely. As a result, much better beam smoothing is expected to be obtained than that of the simulation with same spatial dispersion width. Fig. 2(d)(e) show the experimental result of the beam smoothing effect with the AFGC configuration of a Ti:sapphire PW laser [15], where a strip of paper with 2 mm width is located directly before the FGC compressor which is used to induce diffraction to the laser beam. Fig. 2(d) shows the output laser beam with obviously serious spatial intensity modulation when the compressor is a typical symmetrical FGC. Fig. 2(e) shows the output laser beam with clear smoothed spatial intensity in comparison to that of Fig. 2(d), where only about 90 mm length difference is induced between the G1&G2 and G3&G4 grating pairs. When considering about 100 nm full spectral bandwidth, the induced spatial dispersion width at output of G4 is about 9 mm.

2.2 Wavefront aberrations smoothing owing to spatial dispersion

Actually, in high peak power laser system with big size, besides the relative high spatial intensity modulation at middling/high frequency in the laser beam which is a main damage risk for optics, some of the wavefront aberrations at middling/high spatial frequency will also induce small-scale beam focusing in the far field during free propagation. These wavefront aberrations usually induced by the non-perfect optics, especially for optics with big size. As for the wavefront distortion at low spatial frequency, which will affect the final focal diameter and then the focal intensity, deformable mirrors are usually used to correct these wavefront distortion with low spatial frequency to achieve a small focal diameter. As for the wavefront aberrations with high spatial frequency, spatial filter based on 4f optical system with a small pinhole at the focal point is usually used to filter out the wavefront aberration laser with high spatial frequency. This spatial filter can also be used to magnify the laser beam, which can transmit the nice wavefront from the object plane to the imaging plane of the 4f optical system. To make sure most of the laser beam can pass through the pinhole easily, the size of the pinhole is usually set to be about ten times of its diffraction limited size. As a result, there are still some wavefront aberrations with high spatial frequency and most of the wavefront aberrations with middling spatial frequency left in the output laser beam, which may self-focusing during free propagating to the far field and damage the important optics.

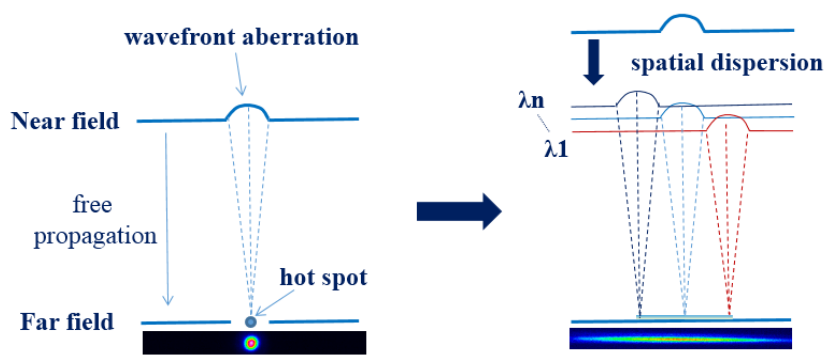

Fig.3 Schematic diagram of wavefront aberration smoothing. Left: near field laser beam with wavefront aberration without spatial dispersion induce hot spot in the far field; Right: the hot spot is smoothed into a line with one dimensional spatial dispersion.

With the induced spatial dispersion, the rest of wavefront aberrations at middling/high spatial frequency cannot be removed, but the self-focusing induced damage risk in the far field can be reduced greatly. Similar to the principle of spatial intensity smoothing [13], the principle of the wavefront aberrations smoothing with spatial dispersion can be explained by using Fig. 3. Without spatial dispersion, the wavefront aberration will induce a hot spot in the far field during free propagating, as shown in the left figure in Fig. 3. Even if one dimensional suitable spatial dispersion width $d$ is induced to the laser beam, the primary hot spot is smoothed into a line for different wavelengths with the spatial dispersion width D, as shown in the right figure in Fig. 3. Since the diameter of hot spot is usually less than 1mm, then even for a 10 mm spatial dispersion width, the hot spot will be reduced by ten times. Even smoothing wavefront can be obtained if the laser beam is smoothed on both X and Y axes. From a different view, the spatial dispersion is actually smooth the wavefront aberrations with middling/high spatial frequency because the wavefront at every position is an average wavefront of all wavelengths, which is the same as the intensity smoothing [13].

## 3. Multistage smoothing based MPC

In our previous work of a typical MPC method, the spatial dispersion is only induced in the pre-compressor stage by using a prism pair [13]. The prism pair induced beam smoothing will protect the first grating directly away from damage risk due to spatial intensity modulations. Beam smoothing in the pre-compressor will also protect the last grating indirectly. Since it is indirectly beam smoothing, some strong spatial intensity modulation or wavefront aberrations at middling/high spatial frequency before, on or after the first grating may transferred onto the last grating which may still damage the last grating.

As for single AFGC, it is very simple. However, the same as previous typical MPC method, single AFGC can only protect the last grating directly [14]. Although, the first grating owns high damage threshold owing to the relative long nanosecond pulse duration, hot spots or other induced strong spatial intensity modulation may also damage the first grating sometimes. In comparison to typical MPC, without the induced spatial dispersion before the FGC, there is no additional input/output pulse energy from the induced spatial dispersion with anti-phase. Then, the maximum input/output laser will be a little bit smaller than that with some anti-phase spatial dispersion before the FGC. Furthermore, in some PW laser system, the main and the biggest amplification crystal also stand for damage risk.

To solve the above problems appeared in single typical MPC or single AFGC, here, two beam smoothing optical designs are combined together in one compressor. Then, beam smoothing is not induced in single stage but is separated into multiple stages, which is named as multistage smoothing based MPC (MS-MPC). The MS-MPC can be separated into two kinds of optical setup. Firstly, the input is a laser beam with no spatial dispersion, the spatial dispersion width is increased step by step in every stage. The largest spatial dispersion width and the smoothest beam is obtained in the compressed output laser beam, which is named as forward MS-MPC. The second one is that the input laser beam owns the largest spatial dispersion, while the spatial dispersion width is decreased step by step in every stage. At last, output laser beam with the smallest or even zero spatial dispersion width is achieved, which is named as backward MS-MPC.

As for the backward MS-MPC, the principle scheme is shown in Fig. 4(a). It will induce complicated static and dynamic spatiotemporal aberrations during relatively long optical propagation and amplification processes, which will affect the output focal spatiotemporal properties. The temporal contrast may also be decreased due to spatiotemporal errors in many imperfect optics that is very important for ultrahigh peak power laser pulses [16-18]. Furthermore, it is not so easy to obtain the best amplification using a beam with large spatial dispersion on both sides, the amplification efficiency will be decreased due to narrowed chirped pulse duration on spatial dispersion region. Then, we will only discuss the first kind of forward MS-MPC in detail here.

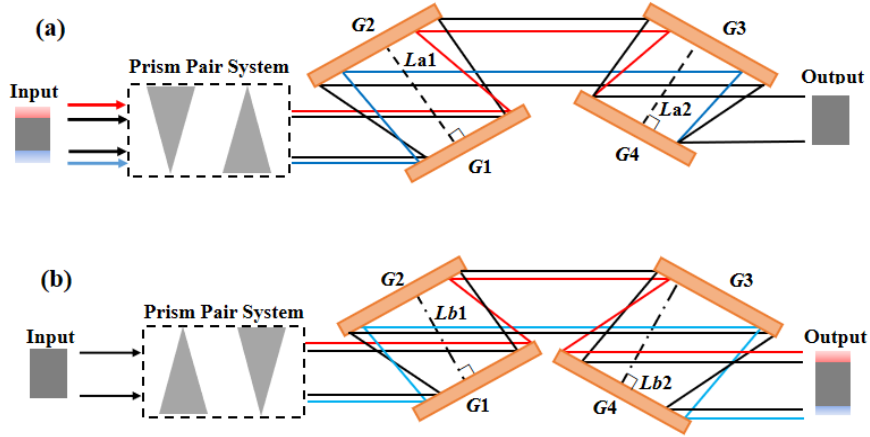

Fig. 4. Principle scheme of (a) backward MS-MPC, with La1>La2, and (b) forward MS-MPC with Lb2>Lb1, where the post-compressor based on spatiotemporal focusing is not shown.

As for the forward MS-MPC shown in Fig.4 (b), suitable spatial dispersion will be induced in the pre-compressor by using one or two prism pairs with relatively short distance and prisms with small apex angle. According to the simulation result, the spatial intensity modulation will be reduced rapidly from 2 times to about 1.3 times with relative short distance of prism pair [13]. Since the first grating owns high damage threshold because of the relative long nanosecond pulse duration, the beam smoothing owing to the induced relative small spatial dispersion is far more enough to protect the first grating from damage directly. Furthermore, when inducing spatial dispersion in two perpendicular directions of laser beam, the spot will be extended into an area. In comparison to a spatial dispersion line induced in one direction, the beam smoothing effect is more effective with spatial dispersion induced in two directions. For example, inducing 8 mm or 5 mm spatial dispersion width on both directions equal to inducing 8× 8=64 mm or 5×5=25 mm spatial dispersion width in one direction, respectively. The relative small spatial dispersion width requirement means it is possible to smooth the laser beam well using prism pair with small apex angle of 15 degrees and relative short distance of about 10 m or 6 m, respectively. To shield the compressor from all the random hot spots before the compressor, at least one prism pair should be located directly before the grating-based compressor in principle. If there is a combined beam expanding and relay imaging system, the prism pair may be located before the relay imaging system, which will imaging the near field well smoothed laser beam directly onto the first grating.

Then, the AFGC is used as the main compressor which will induce suitable amount of additional spatial dispersion on the last grating. There is no doubt, this directly induced spatial dispersion will protect the last grating from possibility of damage risk due to random spatial intensity modulations. Furthermore, since the last grating bears the biggest damage risk because of the shortest femtosecond pulse duration, the induced additional spatial dispersion by the AFGC will further smooth the laser beam, and then reduce the damage risk of the last grating. Here, after beam smoothing in both X and Y directions in the pre-compressor stage, only random hot spots with high spatial frequency need to be considered, then relative small spatial dispersion width is enough which corresponding to a small distance difference between two grating pairs. Note that the AFGC will induce same amount of spectral dispersion to the compressed laser pulse in comparison to the typical symmetric FGC if the sum distances of two prism pairs are equal. As a result, the spectral dispersion can be well compensated directly by using the AFGC. At last, the post-compressor based on spatiotemporal self-focusing effect will be used to compensate the induced spatial dispersion from both the prism pair in the pre-compressor and the asymmetric grating pair in the main compressor. The basic optical setup of the forward MS-MPC is shown in Fig. 4(b). Owing to the smoothed beam profile, self-compression process in bulk medium plates using negatively chirped pulses can also be added in the system to shorten the output pulse duration in the near future [19].

Except for the compressor, the main amplifier may also stand for crystal damage risk due to the pump beam with strong spatial intensity modulation. Then, additional beam smoothing stage based on either prism pairs or grating pairs can be added before the main amplifier to smooth the incident laser beam in the main amplifier. From our recent simulation of beam smoothing with two prism pairs on both X and Y axis, the spatial intensity modulation can be reduced from 2.0 to less than 1.1 when about 10 mm spatial dispersion width are induced on both X and Y directions of the input laser beam [20]. Since the laser beam in the main amplifier of 10s-100s PW laser systems should be more than 200 mm in diameter or side-length. Then, the induced about 5 to 10 mm spatial dispersion width even can be cut off from the laser beam after beam smoothing, which will not affect the main amplification. For example, as for a laser beam before the main amplifier with 370 mm×370 mm beam size, the induced energy loss is only about 4% if 8 mm spatial dispersion width are induced on both X and Y directions, which is definitely acceptable before the main amplifier and will almost not affect the output amplified pulse energy. At the same time, this beam smoothing will improve the spatial intensity modulation of the amplified beam, and then protect the biggest and most expensive crystal away from laser induced damage. Note that the beam smoothing at this stage is optional according to the requirement and condition of every whole laser system.

## 4. An example design for 10s-100s PW laser

So far, the central wavelengths of the proposed 10s to 100s femtosecond PW laser facilities are mainly located around 800 nm or 925 nm. The specific 925 nm wavelength is chosen because only DKDP crystal can be grown large enough to support high enough pulse energy for 10s-100s PW currently. At the same time, the widely studied high energy pump pulse is located around 532 nm. Considering these two preinstalled factors, the phase matching conditions of the OPCPA in DKDP result in the around 925 nm central wavelength with a broad bandwidth. The 200 nm ultrabroad spectral gain bandwidth will also benefit to the beam smoothing through inducing spatial dispersion by using prism pair or grating pair.

The optical diagrammatic sketch of an example design for 10s-100s PW laser system with the forward MS-MPC method is shown in Fig.5. To protect the crystal in the main amplifier and the gratings in the FGC, several stages of beam smoothing are added in the traditional PW system. Two prism pairs PS_I with 15 degrees of apex angle are added before the main amplifier, and each prism pair separated by about 10 m. Then, the 378 mm×378 mm laser beam with no spatial dispersion is expand to about 386 mm×386 mm with about 8 mm spatial dispersion width on four edges. After filtering out the central region with full spectral bandwidth using a soft aperture, the incident laser beam before the main amplifier is about 370 mm×370 mm with smoothing beam profile and no spatial dispersion. This smoothing beam will protect the amplification crystal away from damage due to spatial intensity modulation come from front amplification stages. Note that this stage_0 of beam smoothing is optional based on the requirement and condition of every laser system.

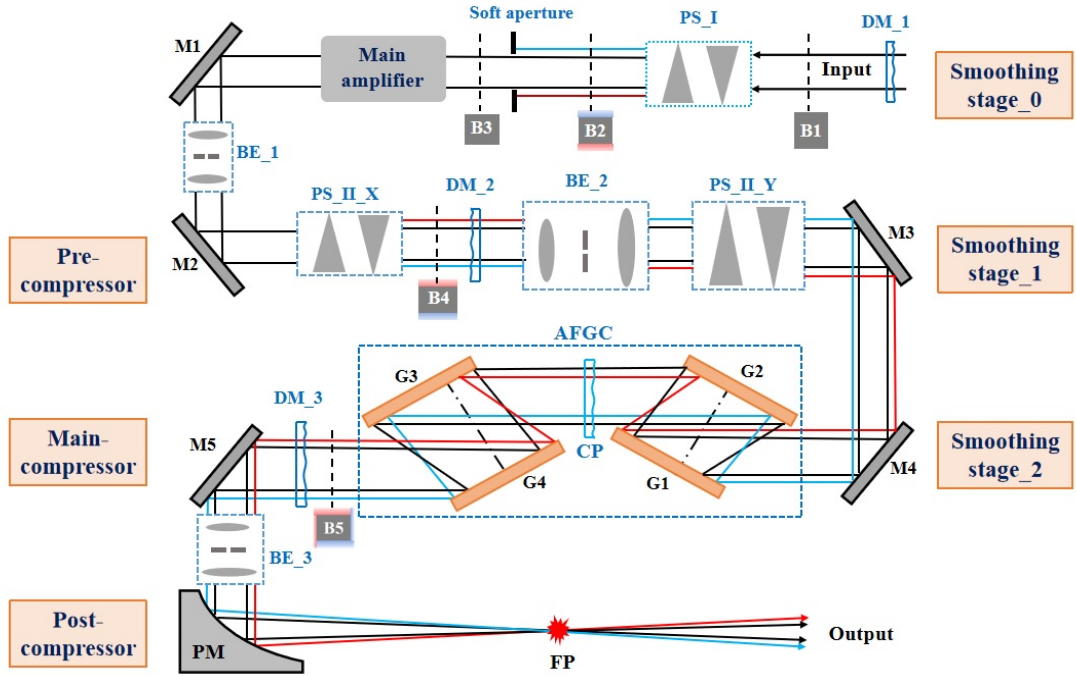

Fig. 5. Optical sketch of a feasible design for 10s-100s PW laser with the forward MS-MPC method. PS_I, PS_II_X, and PS_II_Y: prism pair system for beam smoothing DM_1, DM_2, and DM_3: reflective deformable mirror for wavefront compensation at different stages. BE_1 and BE_2: combined beam expander and relay imaging systems based on two parabolic reflective mirrors. AFGC: asymmetric four-grating compressor with G1-G4 four gratings. M1-M5: plane reflective mirrors. CP: compensate plate. PM: parabolic reflective mirror. FP: focal point. B1-B5: simplified two-dimensional beam profiles.

Anyway, the laser beam after the amplifier and before the forward MS-MPC is assumed to be 370 mm×370 mm with $10^{th}$ order super-Gaussian profiles in both X and Y directions. The laser pulse is a positively chirped pulse with about 4 ns pulse duration, of which the transform limited pulse duration is about 14.5 fs, the center wavelength is 925 nm with about 200 nm spectral bandwidth. After the first 1:1 relay imaging 4f optical system BE_1, the wavefront aberration with high spatial frequency will be filtered out owing to the pinhole in BE_1. Furthermore, the relative small first reflective deformable mirror DM_1 is used to correct the wavefront distortion with low spatial frequency before PS_II_X. As a result, the wavefront before PS_II_X is a well compensated one with only residual aberration in the middling spatial frequency. About 10 mm spatial dispersion width is introduced to the laser beam on the X axis by using a prism pair PS_II_X with 15 degrees apex angle and separated by about 12 m. Then, smoothed laser beam with about 380 mm×370 mm is obtained. Meanwhile, the wavefront aberration with middling spatial frequency is also smoothed owing to the prism pair. Then, another relative big reflective deformable mirror DM_2 is used to correct the wavefront distortion of the optics from M2 to G1. The smoothed laser beam will also help to protect following big optics. Then, the second combined beam expander and relay imaging system BE_2 is used expand the laser beam by about 1.73 times on both sides to about 657 mm×640 mm. Meanwhile the smoothed laser beam together with perfect wavefront after PS_II_X is imaged onto the first grating G1. Then, about 5 mm spatial dispersion width is induced to the laser beam on the Y axis by using another big prism pair PS_II_Y with same 15 degrees apex angle which will further smooth the laser beam. The laser beam is then changed to about 657 mm ×645 mm, where the center about 623 mm×635 mm region owns full spectral bandwidth. Since the prism pair PS_II_Y is located directly before the AFGC, spatial dispersion in Y direction is induced directly to the beam on G1, which will shield all the fixed or random hot spots before the AFGC. As a result, it builds up the first safeguard for all the gratings in the AFGC.

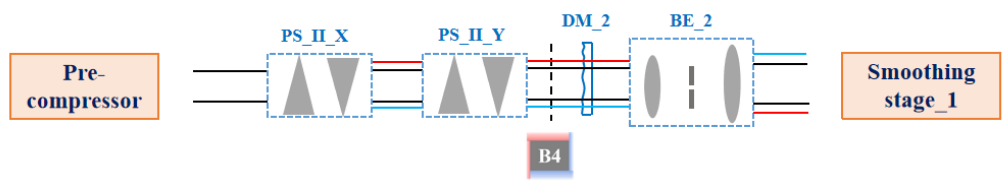

Fig. 6. Another optical setup for smoothing stage_1 in the pre-compressor, where the prism pair PS_II_Y is located before the BE_2 with a relative smaller size and lower cost.

Since the first grating G1 own relative high damage threshold because of nanosecond chirped pulse, which is about 2-3 times higher than that of on the last grating with femtosecond pulse, then the bearable spatial intensity modulation of G1 should be 1.6 times higher than that on the last grating even considering about 60% total compression efficiency. Even if about 1.3 times fluence of damage threshold is reserved for the last grating, then the 1.6 times is increased to more than 2.0 times which is a typical design value for most PW laser system. As a result, the prism pair PS_II_Y can actually be located before the beam expander BE_2 with a relative smaller size and lower cost, as shown in Fig. 6. Here, the output smoothed laser beam after two prism pairs, PS_II_X and PS_II_Y, can be imaged onto the first grating G1 by using beam expander BE_2, which will also owing a smoothed near filed profile. In this case, the prism pair system can only remove the fixed and random hot spots before the beam expanding system BE_2. According to above analysis, G1 have already saved enough above 2.0 times spatial intensity modulation to avoid damage. Furthermore, the spatial intensity modulation induced by the beam expander BE_2 can be removed by the AFGC followed. With this configuration, the laser beam before G1 is changed to about 657 mm×649 mm, where the center about 623 mm×631 mm region owns full spectral bandwidth.

Except for usage as beam smoothing, the two prism pairs with small apex angle may also be used to precisely correct the spatiotemporal aberration, such as pulse front tilt, of the whole PW laser system. In PW laser facilities, to improve the temporal contrast of the output laser pulse, all the transmitted plane optics, such as crystals in the amplifier, are intentionally induced a wedge with tiny angle, which will induce small angular dispersion and pulse front tilt to the output laser beam. In principle, the grating pair in AFGC can be used to compensate the induced angular dispersion by the tiny misalignment the parallel of grating pairs. In a real 10s-100s PW laser system, both the laser beam size and the grating are very large. Then, the requirement of parallel tuning accuracy of the grating pair is very high according to recent reports [11, 17]. As a result, the prism pair can be used to precisely compensate both the induced angular dispersion by transmitted optics and the misalignment the parallel of grating pairs, which will improve the final focal intensity in principle. By the way, the weak reflected beam from the surface of the prisms can also be used to monitor the spectrum, pointing stability, beam profile, time jitter, and wavefront of the laser beam.

After the pre-compressor, the spatiotemporal modified laser beam is then guided into an AFGC with four same 1400 ln/mm gold-coated gratings. The incident angle on the first grating is set to about 61 degrees, which will induce the beam size on the first grating in the diffraction direction is about 657 mm/cos(61°)=1355 mm. The perpendicular distance of the first grating pair is set to 1215 mm, while the second grating pair is set to 1285 mm. Then, there are 70 mm length difference between the two grating pairs, which will induce about 25 mm spatial dispersion width on the last grating with the beam size of 1355+25=1380 mm. The about 1380 mm×650 mm beam size on the last grating is still smaller than that of the currently available biggest grating with a size of about 1450 mm×700 mm. With this 25 mm spatial dispersion width, the possible spatial intensity modulation with high spatial frequency in the laser beam induced by the optics after the last prism pair PS_II_Y can be smoothed very well according to previous simulation, in which a 25 mm spatial dispersion width can reduce the spatial intensity modulation from 2.0 to about 1.2. The experiment results shown in Fig. 2(d) (e) have already shown that even a less than 20 mm spatial dispersion width on G4 can induce very well spatial intensity smoothing. Except for the hot spots with high spatial frequency, when considering the pre-compressor have already induced about 10 mm×5 mm in X and Y directions that equal to 50 mm spatial dispersion width in one direction, the spatial intensity modulation with less than 3 mm period, which according to middling spatial frequency, can also be well smoothed. Then extremely smoothing laser beam can be achieved at the output of the AFGC. Note that the grating pair length difference between the two grating pairs can be conveniently varied according to different requirements, as a result, different spatial dispersion width can be induced by using the AFGC.

Since the total induced spatial dispersion width on the G2 from single spot on G1 is about 440 mm, it needs to be noted that there is about 170 mm width light cutting on the second grating G2 on both edges. The cutting of light by the sharp edges of the grating will induce diffraction and then spatial intensity modulation to the laser beam after the grating G2. According to the theory of diffraction at a straight-edge [21], the diffracted light intensity varies with the distance between the diffracted light source and the receiving screen, where the maximum diffraction intensity can reach about 1.5 times of the initial intensity. Based on all above input parameters, the red dotted line of Fig. 7 shows the one-dimensional beam profile after G2 without light cutting on both sides. In the AFGC, the distance between G2 and G3 is different for two edges. Here, in our diffraction calculation, the shortest distance of the upper edge with long wavelengths is set to 3 m, while the longest distance of the bottom edge with short wavelength is set to 5.5 m. The one-dimensional beam intensity profile on the grating G3 can be obtained by sum of all the diffraction profiles at different wavelengths, as shown in the blue line of Fig. 7. As we can see obviously that the intensity of the light cutting on both edges is about one fifth that of the maximum of the laser beam. From the calculation, the induced spatial intensity modulation by the edge diffraction will not exceed and affect the maximum top-hat part. And then it will not induce damage risk to the G3. The induce energy loss by the both edges cutting is calculated to be about 1.9%, which is absolutely acceptable for an AFGC system. Note that the diffraction induce spatial intensity modulation by G2 in the laser beam is absolutely smoothed after G4 owing to the induced large spatial dispersion by the grating pair G3 and G4. Actually, spectral cutting on G2 had been widely used in many grating based compressors. The program for the edge diffraction used here is experimentally proved in a kHz Ti:sapphire laser system with about 15 mm beam diameter, as shown in Fig. 7(b-e). The simulation results of edge diffraction are Fig.7 (b) and (d), the experimental results are Fig.7 (c) and (e), where red line is one dimensional beam intensity profile without spectral cutting and blue line is one dimensional beam intensity profile with spectral dispersion width cutting on both sides. The middle bottom insets of Fig.7 (c) and (e) are two dimensional beam intensity profile, and (c) and (e) are spectral dispersion width cutting at the intensity of about 0.2 and 0.5 that of the highest value on both sides, respectively.

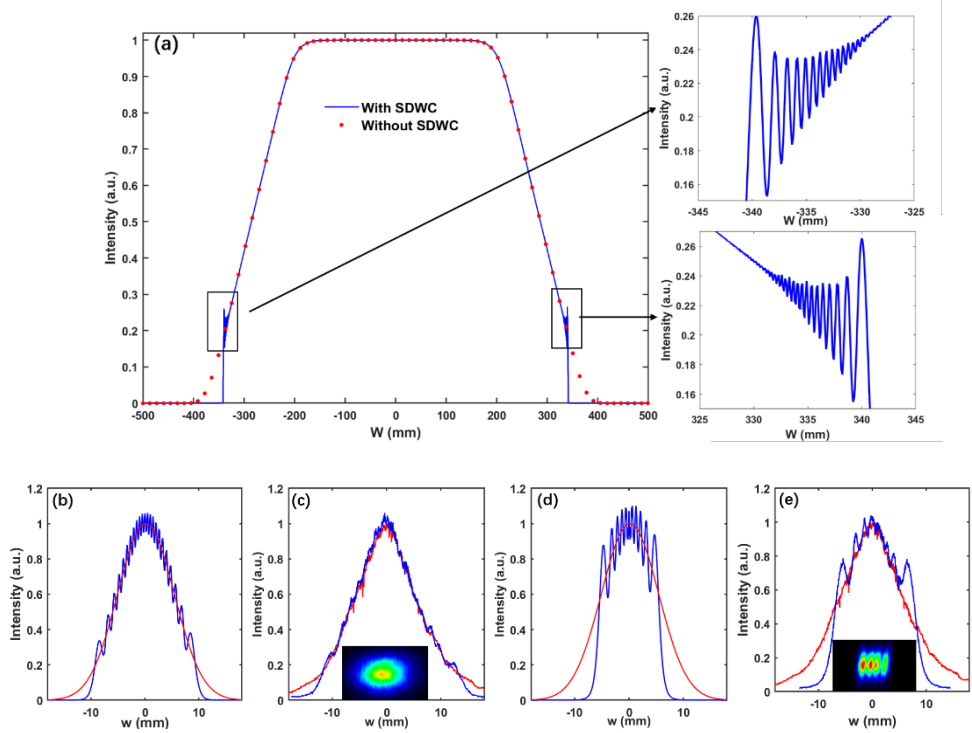

Fig.7. (a) The one dimensional beam profile before G3 without spectral cutting (red dotted line ) and with about 150 mm spectral dispersion width cutting on both sides (blue solid line), where SDWC: spectral dispersion width cutting. The insets on the right are the calculated diffraction induced modulation for both edges cutting. (b) and (d) the edge diffraction of simulation results, (c) and (e) the edge diffraction of experimental results

As the full spectral bandwidth regions on each grating will suffer the strongest laser fluence, then we only need to consider the laser fluence of this region with full spectral bandwidth. As for the first grating, the same as previous work on typical MPC, the effective beam size on the grating is about (623+(657-623)/2)/cos(61º)=1320 mm in the X direction, and 635+(645-635)/2=640 mm or 631+(649-631)/2=640 mm in the Y direction. For a 2500 J input amplified laser pulse, the laser fluence is 2500/(132×64)≈296 mJ/cm$^2$, which is 2 times lower than the grating damage threshold about 600 mJ/cm$^2$ at nanosecond pulse duration [13, 22]. In the AFGC, laser fluence of the region with full spectral bandwidth is decreased to about 296×0.60=178 mJ/cm$^2$ if considering about 60 % total compression efficiency, which is about 1.28 times lower than the damage threshold about 229 mJ/cm$^2$ of a compressed femtosecond laser pulse [13, 22]. The output about 2500×0.60= 1500 J compressed output pulse energy can support 100 PW laser with 15 fs compressed pulse duration. Since the spatial intensity modulation of the output laser beam can be smoothed to less than 1.1 theoretically after three stages of beam smoothing processes, it is possible to achieve 100 PW theoretically by using this MS-MPC with single beam and currently available gold-coated grating. Of course, about 75 PW would be very safe and feasible for a real 10s-100s PW laser system because the laser fluence on the last grating is almost about 1.70 times lower than that of the damage threshold for femtosecond pulse, where the beam spatial intensity modulation is smoothed to less than 1.2. Note that we only show the capabilities of this MS-MPC method for 10s to 100s PW compression, the proposed parameters here are not the best optimized values, of which enough margins have already been saved on many parameters such as the beam size.

In the AFGC, a compensate plate is proposed to compensate the wavefront distortion with low spatial frequency from the two gratings [11], G2 and G3, which will induce serious spatiotemporal coupling error otherwise and decrease the final focal intensity obviously. After the AFGC, the compressed output beam size is about 670 mm×650 mm with the full spectral bandwidth region of about 612 mm×631 mm or 612 mm×635 mm. Then, a very big reflective deformable mirror DM_3 is located directly after the AFGC to compensate the wavefront distortion induced by the optics from G4 to the final reflective parabolic mirror PM. Due to limited number of poles on DM_2 and DM_3, the deformable mirror can only compensate wavefront with low spatial frequency of tens millimeters. Except for beam smoothing, the induced spatial dispersion will also diffuse the wavefront aberration at middling and high spatial frequency regions, which will further help to reduce the spatial intensity modulation out of focal point. This smoothed laser beam can be used to induce a further pulse self-compression in a glass plate with negative chirped input after the final parabolic mirror PM in the future [13, 19].

After the main-compressor with the AFGC setup, the same as the typical MPC, the output laser beam with smoothed spatial intensity and diffused wavefront aberration is expanded and imaged by using BE_3 onto the final reflective parabolic mirror PM. The PM is used as the spatial dispersion compensate mirror. At the focal point, the spatial dispersion will be compensated automatically owing to spatiotemporal focusing and achieved the highest peak power and focal intensity. Note that the damage threshold of the optics after the AFGC can be several times higher than that of gratings. This is because the grating with broad diffraction spectral bandwidth can only be coated with gold which is the weakest film, while the reflective optics after the AFGC can be coated by other kinds of coating film with several times higher damage fluence threshold.

## 5. Discussion and conclusion

Due to the fixed and random hot spots induced by diffraction from dust, defects on optics, and also the spatial uniform pump beam, the spatial intensity modulation of the amplified laser in a PW laser system is usually very high. To avoid laser-induced damage of the important optics, especially gratings, the laser fluence on the gratings usually have to be kept under half that of the grating damage threshold, which undoubtedly affect the maximum output pulse energy of a compressor. Our previous optical designs, the typical MPC and the AFGC based two-step MPC have been successfully improve the output pulse energy by 2-4 times theoretically in comparison to a traditional FGC, 10s to 100s PW lasers can be achieved in a single beam based on both methods [13, 14].

In a real PW laser system, except for the capability to achieve high peak power, the operating safety is also very important to maintain the laser system economically and increase the experimental efficiency. Here, the MS-MPC that combines both the typical MPC and the AFGC optical designs in one compressor system. Both the spatial intensity modulation and wavefront aberrations with middling/high spatial frequency can be smoothed by using the induced spatial dispersion through the grating pair or the prism pair. Then, the new MS-MPC design can improve the operating safety of PW laser systems greatly. An AFGC is used as the main compressor which will induce spatial dispersion on the last grating and protect the last and weakest grating directly. In comparison to typical MPC, the induced spatial dispersion width by the prism pair which located directly before the grating-based compressor can be small, otherwise, it needs very long distance between two prisms to induce relative large spatial dispersion width. Moreover, in MS-MPC, an additional beam smoothing stage_0 can be added before the main amplifier to protect the biggest and expensive amplification crystal away from damage.

In conclusion, an improved MPC named MS-MPC is proposed to achieve 10s-100s PW laser with single laser beam. At the same time, the proposed MS-MPC can protect the important optics such as the biggest crystal, all four gold-coated gratings directly from laser induced damage due to hot spots Since the distance of different prism pairs or grating pairs can be varied easily, compressed output laser beam with different spatial dispersion width can be achieved conveniently which may induce different laser fields around the focal point and benefit the laser physics experiments. This MS-MPC optical design can be used easily in all exist PW laser facilities to improve their output pulse energy and running safety.

**Funding.** This work is supported by the National Natural Science Foundation of China (NSFC) (61527821, 61905257, U1930115), and Shanghai Municipal Science and Technology Major Project (2017SHZDZX02).

**Acknowledgments.** The authors would like to thank Dr. Chenqiang Zhao, Dr. Cheng Wang, and Dr. Yi Xu for the help on gratings and PW laser.

**Disclosures.** The authors declare no conflicts of interest.

**References**

1. A. Dubietis, G. Jonusauskas, and A. Piskarskas, "Powerful femtosecond pulse generation by chirped and stretched pulse parametric amplification in BBO crystal," Optics Communications **88**, 437-440 (1992).
2. D. Strickland and G. Mourou, "Compression of Amplified Chirped Optical Pulses," Optics Communications **55**, 219-221 (1985).
3. C. N. Danson, C. Haefner, J. Bromage, T. Butcher, J. Chanteloup, E. A. Chowdhury, A. Galvanauskas, L. A. Gizzi, J. Hein, and D. I. Hillier, "Petawatt and exawatt class lasers worldwide," High Power Laser Science and Engineering **7**, e54 (2019).
4. W. Q. Li, Z. B. Gan, L. H. Yu, C. Wang, Y. Q. Liu, Z. Guo, L. Xu, M. Xu, Y. Hang, Y. Xu, J. Y. Wang, P. Huang, H. Cao, B. Yao, X. B. Zhang, L. R. Chen, Y. H. Tang, S. Li, X. Y. Liu, S. M. Li, M. Z. He, D. J. Yin, X. Y. Liang, Y. X. Leng, R. X. Li, and Z. Z. Xu, "339 J high-energy Ti:sapphire chirped-pulse amplifier for 10 PW laser facility," Optics Letters **43**, 5681-5684 (2018).
5. F. Lureau, G. Matras, O. Chalus, C. Derycke, T. Morbieu, C. Radier, O. Casagrande, S. Laux, S. Ricaud, G. Rey, A. Pellegrina, C. Richard, L. Boudjemaa, C. Simon-Boisson, A. Baleanu, R. Banici, A. Gradinariu, C. Caldararu, B. De Boisdeffre, P. Ghenuche, A. Naziru, G. Kolliopoulos, L. Neagu, R. Dabu, I. Dancus, and D. Ursescu, "High-energy hybrid femtosecond laser system demonstrating 2 x 10 PW capability," High Power Laser Science and Engineering **8**, e43 (2020).
6. J. D. Zuegel, S. W. Bahk, I. A. Begishev, J. Bromage, C. Dorrer, A. V. Okishev, J. B. Oliver, and Ieee, "Status of High-Energy OPCPA at LLE and Future Prospects," in *Conference on Lasers and Electro-Optics (CLEO)*, Conference on Lasers and Electro-Optics 2014), 2 pp(2014).
7. A. Shaykin, I. Kostyukov, A. Sergeev, and E. Khazanov, "Prospects of PEARL 10 and XCELS Laser Facilities," Rev. Laser Eng. (Japan) **42**, 141-144 (2014).
8. E. Cartlidge, "Eastern Europe's laser centers will debut without a star," Science **355**, 785 (2017).
9. Cartlidge and Edwin, "The light fantastic," Science **359**, 382-385 (2018).
10. N. Blanchot, G. Marre, J. Neauport, E. Sibe, C. Rouyer, S. Montant, A. Cotel, C. Le Blanc, and C. Sauteret, "Synthetic aperture compression scheme for a multipetawatt high-energy laser," Applied Optics **45**, 6013-6021 (2006).
11. J. Liu, X. Shen, Z. Si, C. Wang, C. Zhao, X. Liang, Y. Leng, and R. Li, "In-house beam-splitting pulse compressor with compensated spatiotemporal coupling for high-energy petawatt lasers," Optics Express **28**, 22978-22991 (2020).

12. G. Zhu, J. V. Howe, M. Durst, W. Zipfel, and C. Xu, "Simultaneous spatial and temporal focusing of femtosecond pulses," Optics Express **13**, 2153-2159 (2005).
13. J. Liu, X. Shen, S. M. Du, and R. X. Li, "Multistep pulse compressor for 10s to 100s PW lasers," Optics Express **29**, 17140-17158 (2021).
14. X. Shen, S.Du, J. Liu, and R. Li, "Asymmetric four-grating compressor for ultrafast high power lasers," arXiv:2105.04863 [physics.optics] (2021).
15. Z. X. Zhang, F. X. Wu, J. B. Hu, X. J. Yang, J. Y. Gui, P. H. Ji, Y. Q. Liu, C. Wang, Y.Q.Liu, X. M. Lu, Y. Xu, Y. X. Leng, R. X. Li, and Z. Z. Xu, "The 1PW/0.1Hz laser beamline in SULF facility," High Power Laser Science and Engineering, 8, E4 (2020).
16. Z. Li, K. Tsubakimoto, H. Yoshida, Y. Nakata, and N. Miyanaga, "Degradation of femtosecond petawatt laser beams: Spatio-temporal/spectral coupling induced by wavefront errors of compression gratings," Applied Physics Express **10**, 102702 (2017).
17. Z. Y. Li and N. Miyanaga, "Simulating ultra-intense femtosecond lasers in the 3-dimensional space-time domain," Optics Express **26**, 8453-8469 (2018).
18. P. Wang, X. Shen, J. Liu, and R. Li, "Single-shot fourth-order autocorrelator," Advanced Photonics **1**(2019).
19. J. Liu, X. W. Chen, J. S. Liu, Y. Zhu, Y. X. Leng, J. Dai, R. X. Li, and Z. Z. Xu, "Spectrum reshaping and pulse self-compression in normally dispersive media with negatively chirped femtosecond pulses," Optics Express **14**, 979-987 (2006).
20. S. Du, X. Shen, W. Liang, P. Wang, and J. Liu, "Beam smoothing based on prism pair for multistep pulse compressor in PW lasers," arXiv:2110.12575 [physics.optics](2021)
21. Y. Ohtsuka and M. C. Yin, "Fresnel diffraction by a semitransparent straight edge object with acoustically coherence-controllable illumination," Applied Optics **23**, 300-306 (1984).
22. P. Poole, S. Trendafilov, G. Shvets, D. Smith, and E. Chowdhury, "Femtosecond laser damage threshold of pulse compression gratings for petawatt scale laser systems," Optics Express **21**, 26341-26351 (2013).